\RequirePackage{fix-cm}
\documentclass[twocolumn]{svjour3}          

\smartqed  
\usepackage{graphicx}
\usepackage{amsmath}
\usepackage{hyperref}

\usepackage{subcaption}
\usepackage{amssymb}
\usepackage{xcolor}
\definecolor{newcolor}{rgb}{.8,.349,.1}

\newcommand{\del}{\boldsymbol{\nabla}}

\newcommand\Ra{\mbox{\textit{Ra}}}  
\newcommand\Pra{\mbox{\textit{Pr}}} 
\newcommand\Nu{\mbox{\textit{Nu}}} 

\usepackage[]{algorithm2e}
\usepackage{pythonhighlight}
\usepackage{placeins}
\usepackage{nth}

\journalname{}
\begin{document}


\title{Performance of parallel-in-time integration for Rayleigh B\'enard Convection\thanks{A.C. was supported by Engineering and Physical Sciences Research Council (EPSRC) Centre for Doctoral Training in Fluid Dynamics (EP/L01615X/1). C.J.D is supported by a Natural Environment Research Council (NERC) Independent Research Fellowship (NE/L011328/1). D.R. thankfully acknowledges partial support by NERC grant NE/R008795/1 "Parallel Paradigms for Numerical Weather Prediction". S.M.T. was supported by funding from the European Research Council (ERC) under the EU's Horizon 2020 research and innovation programme (grant agreement D5S-DLV-786780).
J.S.O. and A.C. were supported by NASA LWS grant no. NNX16AC92G. This work used the ARCHER UK National Supercomputing Service (http://www.archer.ac.uk) as well as ARC2 and ARC3, part of the High Performance Computing facilities at the University of Leeds. Figures were produced using Matplotlib \cite{hunter2007matplotlib}. }}



\author{Andrew Clarke         \and
        Chris Davies    \and
        Daniel Ruprecht \and
        Steven Tobias   \and
        Jeffrey S. Oishi
}


\institute{Andrew Clarke \at
              Centre for Doctoral Training in Fluid Dynamics, University of Leeds, Leeds, United Kingdom. 
              \email{scatc@leeds.ac.uk}           
           \and
           Christopher Davies \at
              School of Earth and Environment, University of Leeds, Leeds, United Kingdom. 
              \email{c.davies@leeds.ac.uk}
            \and
            Daniel Ruprecht \at
                Lehrstuhl Computational Mathematics, Institut für Mathematik, Technische Universität Hamburg, Germany.
                \email{ruprecht@tuhh.de}
            \and
            Steven Tobias \at
                Dept. of Applied Mathematics, University of Leeds, Leeds, United Kingdom.
                \email{s.m.tobias@leeds.ac.uk}
            \and
            Jeffrey S. Oishi \at
                Dept. of Physics \& Astronomy
                Bates College, Lewiston, ME, USA
                \email{joishi@bates.edu}
}

\date{Received: date / Accepted: date}

\maketitle

\begin{abstract}
Rayleigh-B\'enard convection (RBC) is a fundamental problem of fluid dynamics, with many applications to geophysical, astrophysical, and industrial flows. 
Understanding RBC at parameter regimes of interest requires complex physical or numerical experiments. 
Numerical simulations require large amounts of computational resources; in order to more efficiently use the large numbers of processors now available in large high performance computing clusters, novel parallelisation strategies are required. 
To this end, we investigate the performance of the parallel-in-time algorithm Parareal when used in numerical simulations of RBC. 
We present the first parallel-in-time speedups for RBC simulations at finite Prandtl number. 
We also investigate the problem of convergence of Parareal with respect to to statistical numerical quantities, such as the Nusselt number, and discuss the  importance of reliable online stopping criteria in these cases.

\end{abstract}

\section{Introduction}
\label{intro}

Rayleigh-B\'enard convection (RBC) is an archetypal problem in fluid dynamics, describing the buoyancy driven flow of a fluid heated from below and cooled from above \cite{ahlers2009heat}. 
It allows for studying fundamental properties of fluid dynamics and is used as a simplified analogue for astrophysical and geophysical systems such as planetary interiors, stars, and the atmosphere \cite{busse1978non}, \cite{getling1998rayleigh}. 

RBC is the convection of a fluid driven by a vertical temperature gradient $\Delta T$ between two horizontal plates separated by a distance $L$.
The problem can be characterised by three non-dimensional parameters. 
The Rayleigh number is given by
\begin{equation}
    \Ra = \frac{\alpha g \Delta T L^3}{\nu \kappa},
\end{equation}
where $\alpha$ is the coefficient of thermal expansion, $\nu$ is the kinematic viscosity of the fluid, $g$ is gravity, and $\kappa$ is the thermal diffusivity. 
The Prandtl number is
\begin{equation}
    \Pra = \frac{\nu}{\kappa}.
\end{equation}
The third controlling feature of the flow is the aspect ratio of the domain, $L_x/L_z$ where $L_x$, $L_z$ are the horizontal and vertical size of the domain.
The Rayleigh number is a measure of how much the flow is driven by the temperature, while the Prandtl number is an inherent property of the fluid. 

Very high or infinite Prandtl number is used as a model for convection in the Earth's mantle~\cite{samuel2012time}, while a Prandtl number $ \sim 1$ is commonly used in simulations of the Earth's core, see for example \cite{mound_davies_2017,matsui2016performance}. 
In this work we investigate cases where $\Pra =1$, $L_x/L_z = 2$ and focus on the effects of changes in the Rayleigh number.


Rayleigh-B\'enard convection has been studied intensively throughout the last few decades and before, see for example the papers by Siggia \cite{siggia1994high} or Verzicco and Camussi \cite{verzicco2003numerical}. Some notable studies utilising Rayleigh-B\'enard convection include Cattaneo et al. \cite{cattaneo2003interaction} who studied solar magnetic field interactions, Glatzmaier and Roberts \cite{glatzmaier1995three}, who produced the first simulation of a geomagnetic field reversal, and McKenzie et al. \cite{mckenzie1974convection} who studied the effect of mantle flow in the earth.
For more in depth reviews of the subject, see for example Bodenschatz et al. \cite{bodenschatz2000recent} or Ahlers et al. \cite{ahlers2009heat}.

Much interest has developed in the behaviour of a fluid convecting at high Rayleigh numbers. This is an important area of study, as high Rayleigh numbers are thought to be present in many geophysical and astrophysical bodies. 
Different scaling regimes are believed to exist at different orders of Rayleigh number, and much work has been done to find the exact scaling behaviour of the Nusselt number ($\Nu$, defined below), with $Ra$, see for example Grossmann and Lohse \cite{grossmann2000scaling}, Cioni et al.\cite{cioni1997strongly}, Kerr \cite{kerr1996rayleigh}, and Siggia \cite{siggia1994high}.

To test the theories describing this behaviour, experiments at higher and higher $Ra$ are required, a difficult task to achieve, either numerically or experimentally. Much of this work is now done through direct numerical simulations, see for example Zhu et al. \cite{zhu2018transition} and Schumacher \cite{schumacher2016transitional}.
These studies require an enormous amount of computational power~\cite{kooij2018comparison} and, due to constraints on parallel performance, there is a need to investigate further options for increasing the degree of parallelism in simulation codes.

One such option which has gained much interest in recent years is parallel-in-time integration. 
This allows the time domain to be parallelised in a similar way to how the spatial domain is commonly parallelised.
The recent interest in parallel in time methods was sparked by the introduction of the Parareal algorithm by Lions et al. \cite{lions2001resolution}. 
Subsequently, much research has been carried out in this area; new parallel in time algorithms such as Parallel Full Approximation Scheme in Space and Time (PFASST, Minion \cite{minion2010hybrid}), Parallel implicit time-integrator (PITA, Farhat and Chandesris \cite{FarhatEtAl2003}), and Multigrid Reduction in Time (MGRIT, Friedhoff et al. \cite{FriedhoffEtAl2013}) have been proposed. 
For a comprehensive review see for example Gander~\cite{gander201550}. 

In this work, we present the first reported speedup from parallel-in-time integration for the problem of RBC at finite Prandtl number. 
We extend the work of Samuel \cite{samuel2012time}, who studied the performance of Parareal for infinite Prandtl, into a regime with more varied geo- and astro- physical applications. 
For infinite Prandtl number, the time derivative in the momentum equation vanishes and temperature is the only prognostic variable.
In contrast, for finite Prandtl number, both velocity and temperature have to be integrated in time.
Samuel reported speedups of up to 10 were found when using up to 40 CPUs for infinite Prandtl number, when combining Parareal with spatial parallelisation. 
These results were largely in line with the theoretical performance model they developed.
Recently, Kooij \cite{kooij2017towards} discussed parallel-in-time methods as an attractive option for simulations of Rayleigh B\'enard convection, but did not supply any results in this direction.

Our results show that Parareal can faithfully reproduce the relationship between Rayleigh- and Nusselt number found in the literature.
Given that the number of studies of Parareal for problems with non-linear complex dynamics is limited, this is a useful result in itself.
We further investigate the convergence properties of Parareal with respect to the $L^2$ error between individual trajectories as well as averaged quantities.
While the former is typically used as a termination criterion for Parareal, the latter is often more relevant for applications.
Our results show that, particularly for flows at high Rayleigh number, Parareal can fail to converge to the fine trajectory while still converging to the correct averaged dynamics.
Only at Rayleigh numbers beyond $10^7$ does Parareal's convergence start to deteriorate.
This suggests that research into alternative termination criteria for Parareal, aimed at reproducing correct statistics instead of individual trajectories, would be a useful direction for future research.

\section{Rayleigh B\'enard Convection}

\subsection{Equations and Domain}

We use the Boussinesq approximation to the Navier-Stokes equations for fluid flow in a 2D Cartesian domain. The non-dimensional Oberbeck-Boussinesq equations modelling Rayleigh - B\'enard convection can be written as
\begin{equation}
    \frac{1}{\Pra} \left( \frac{\partial \boldsymbol{u}}{\partial t} + \boldsymbol{u} \cdot \del \boldsymbol{u} \right) = -\del p + \Ra T \cdot \boldsymbol{\hat{z}} + \nabla^2 \boldsymbol{u}, 
\end{equation}

\begin{equation}
    \del \cdot \boldsymbol{u} = 0,
\end{equation}

\begin{equation}
    \frac{\partial T}{\partial t} + \boldsymbol{u} \cdot \del T = \nabla^2 T,   
\end{equation}
with fixed temperature 
\begin{equation}
    T|_{z=-0.5} = 1, ~~ T|_{z=0.5} = 0, ~~ \boldsymbol{u}|_{z=-0.5,0.5} = 0,
\end{equation}
and fixed flux
\begin{equation}
     \left.\frac{\partial T}{\partial z}\right|_{z=-0.5,0.5} = -1,  ~~ \boldsymbol{u}|_{z=-0.5,0.5} = 0
\end{equation}
boundary conditions.

Here, $\boldsymbol{u}=(u,w)$ represents the horizontal and vertical velocity of the fluid, $T$ represents the temperature, $t$ represents time and $p$ is pressure. The fundamental time scale is taken as a thermal diffusion time $\tau_d \sim L^2/\kappa$, $T$ is scaled by $\Delta T$, and length is scaled by $L$. We use a domain of size ($x=2,~ z=1$), where $x$ is the horizontal direction, and $z$ the vertical, giving an aspect ratio of 2. 
We begin with a linear temperature profile with small perturbations and $\boldsymbol{u}=0$.

In the fixed flux case, we use the flux Rayleigh number $\Ra_f$, defined as
\begin{equation}
    \Ra_f = \frac{\alpha g \beta L^4}{\nu \kappa},
\end{equation}
where $\beta$ is the imposed vertical heat flux, instead of the standard Rayleigh number in the momentum equation. 
The flux Rayleigh number can be related to the standard Rayleigh number as $\Ra_f=\Ra \Nu$ \cite{johnston2009comparison}.

\subsection{Consistency Checks}
\label{section:consistency_checks}
The Reynolds number can be computed from the velocity of the fluid. 
A characteristic speed $U$ is determined as $\overline {\langle u^2 + w^2 \rangle^{1/2}}$ where the overbar denotes the time average and $\left \langle \cdot \right \rangle$ the volume average. 
Our parameters are chosen such that $Re = U$

The amount of heat transport due to convection is represented by the Nusselt number
\begin{equation}
\label{eq:nusselt_integral}
   \Nu_V =  \frac{1}{V}\int_V \left({  -\frac{\partial T}{\partial z} +wT} \right)  \mathrm{d}V,
\end{equation} 
where the subscript $V$ indicates that it has been calculated using a volume integral over the domain.
A Nusselt number of 1 indicates that all heat transport is due to conduction, whilst $\text{Nusselt}>1$ indicates advection is present. A larger Nusselt number indicates more heat transport by advection. 

In order to confirm the accuracy of our simulations, we carry out three internal consistency checks. We calculate the Nusselt number in three ways. First, integrated over the domain volume via Equation~\ref{eq:nusselt_integral}. Second, on the bottom plate via
\begin{equation}
\label{eq:nu_bottom}
    \Nu_b = \left \langle \left. {-\frac{\partial T}{\partial z} }  \right \rangle_H \right|_{z=-0.5},
\end{equation} 
where $\langle a \rangle_H = L_x^{-1}\int_{x=0}^{x=L_x}{a}~\mathrm{d}x$ is a horizontal plane average.
Third, on the top plate via
\begin{equation}
\label{eq:nu_top}
    \Nu_t = \left \langle \left. -\frac{\partial T}{\partial z}   \right \rangle_H \right|_{z=0.5}.
\end{equation}
Conservation of energy requires 
\begin{equation}
    \Nu = \overline{\Nu}_b = \overline{\Nu}_t = \overline{\Nu}_V,    
\end{equation}
  \cite{king2012heat}. The standard test in the literature is for the Nusselt numbers calculated at different heights of the domain to be within 1\% of each other~\cite{king2012heat,mound_davies_2017,stevens2010radial}.

Thus, we calculate the maximum relative difference between the bulk Nusselt number and the Nusselt numbers at the top, bottom as well as the difference between the top and bottom Nusselt number 
\begin{equation}
\label{eq:nu_int_error}
    {\Nu_\text{int}} = \frac{\max \left( |\overline{\Nu}_b - \overline{\Nu}_V|, |\overline{\Nu}_b-\overline{\Nu}_t|,|\overline{\Nu}_V-\overline{\Nu}_t|   \right)}{\overline{\Nu}_V} .
\end{equation}

As a second consistency check, we verify that buoyancy generation is balanced with viscous dissipation.
If we average over a sufficiently long time, the $\frac{D \boldsymbol{u}}{D t}$ term of the momentum equation goes to zero. We then take the dot product of the momentum equation with $\boldsymbol{u}$ and integrate to find the energy balance
\begin{equation}
    | \overline{\boldsymbol{u} \cdot \nabla^2 \boldsymbol{u}} | = | \overline{\boldsymbol{u} \cdot RaT \hat{\boldsymbol{z}}} |,
\end{equation}
where the first term represents the viscous dissipation $\epsilon_U$, and the second term represents the buoyancy production $P$. The standard test in the literature is for simulations to find these quantities within 1\% of each other \cite{king2012heat,mound_davies_2017}.
We check this by calculating
\begin{equation}
\label{eq:balance}
    \frac{|P - \epsilon_U|}{P}.
\end{equation}

\begin{figure}
    \centering
    \includegraphics{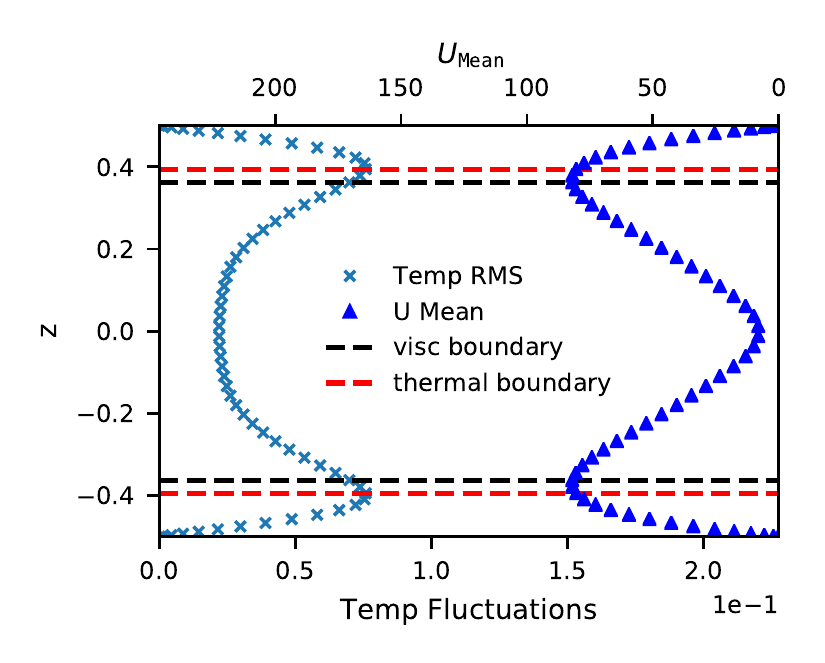}
    \caption{Rayleigh B\'enard flow at Rayleigh number $=10^5$. Temperature fluctuations (left side of graph, bottom scale) denote the $T_\text{rms}$ of the temperature field (defined in text), $U_\text{Mean}$ (right side of graph, top scale) denotes the magnitude of the horizontal component of the velocity. The thermal boundary layer is defined by the height at which the peak $T_\text{rms}$ is found, and the viscous boundary layer is defined by the height at which the peak $U_\text{mean}$ is found \cite{king2013scaling}.  }
    \label{fig:fluctuations}
\end{figure}

As a third test, we make sure that the boundary layers are resolved with a minimum number of nodes.
The thermal boundary layer can be defined using the peak value of $T_{rms}$, calculated as
\begin{equation}
\label{eq:t_rms}
T_\text{rms}(z) = \overline{\left \langle \sqrt{\left(T- \overline{\langle T\rangle_H} \right)^2}   \right \rangle}_H
\end{equation}
as in King et al., \cite{king2013scaling}. Figure \ref{fig:fluctuations} shows the relationship between $T_\text{rms}$ and the thermal boundary layers, and the relationship between the viscous boundaries and the mean horizontal velocity magnitude. The thickness of the thermal boundary layer $\delta_T$ is defined by the height at which the peak value of $T_\text{rms}$ occurs. The boundary layer scales with the Nusselt number as 
\begin{equation}
    \delta_t = \frac{1}{2} L \Nu^{-1},
\end{equation}
see Grossman and Lohse~\cite{grossmann2000scaling}. 
The thermal boundary layer plays a significant role in the properties of Rayleigh-B\'enard convection, and it is essential that they are fully resolved in any numerical simulation~\cite{shishkina2010boundary}. 
Amati et al.~\cite{amati2005turbulent} showed that at least 4 grid points are required in the thermal boundary layer, while Verzicco and Camussi \cite{verzicco2003numerical} stated that 6 cells are needed. 
Stevens et al.~\cite{stevens2010radial} say that up to 7 points could be the minimum number of points required. 
In this work, we specify that at least 6 points are in the boundary layer. 
The number of points in the thermal boundary layer will be denoted as $N_\text{BL}$.

\begin{figure}
    \centering
    \includegraphics{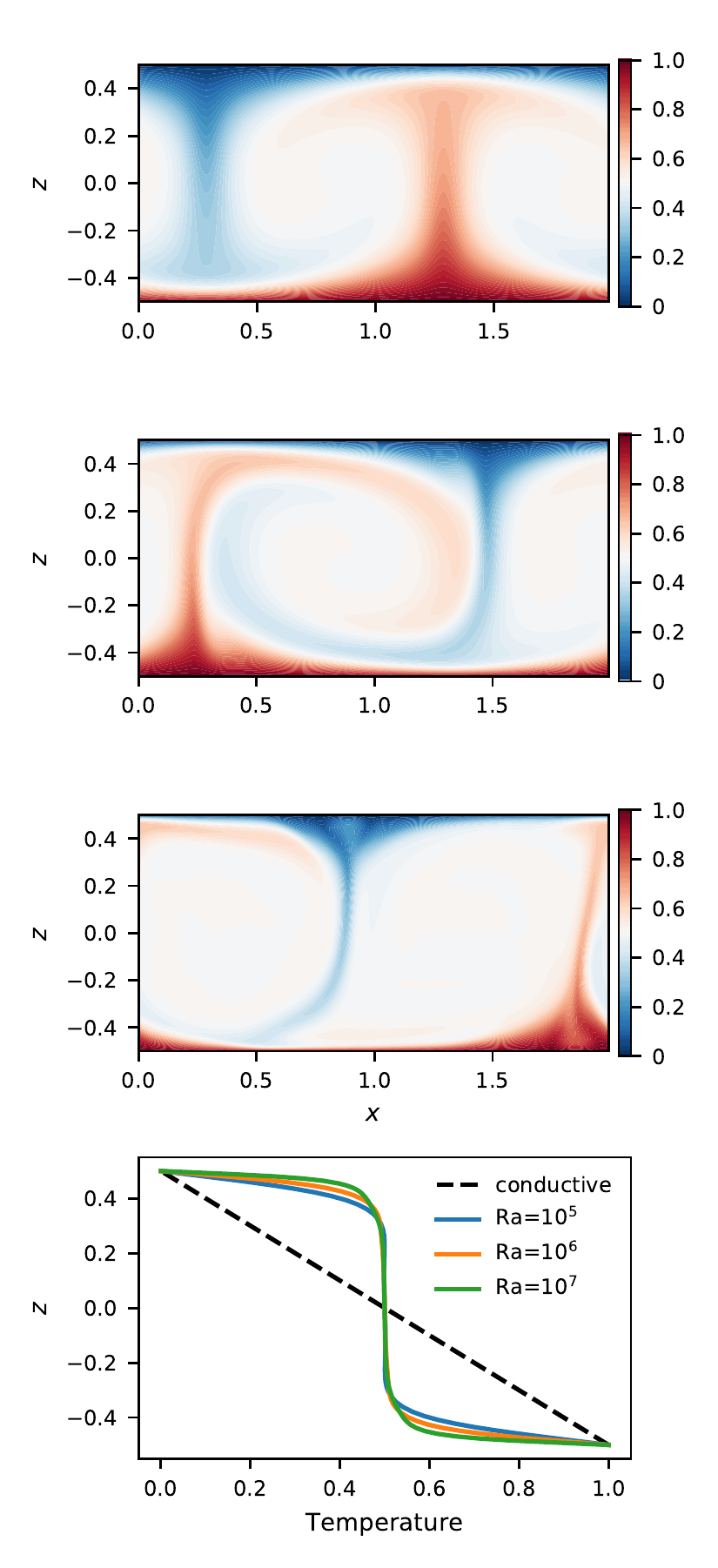}
    \caption{Temperature field for flows with $Ra=10^5$ (top), $10^6$ (middle top), and $10^7$ (middle lower) taken after a statistically steady state has been reached. The bottom plate is fixed at $T=1$, whilst the top plate is fixed at $T=0$, and both top and bottom plates are no-slip. There is steady flow for $Ra=10^5$, with more unsteady and smaller plumes at $10^6$, and even more so at $10^7$. At $Ra=10^7$, there is a small amount of entrainment of fluid into the base of the plumes. The bottom figure shows temperature profiles for all three cases, compared to the purely conductive case. Boundary layers get thinner as $\Ra$ increases. }
    \label{fig:field_snapshot_ra_all}
\end{figure}

Figure~\ref{fig:field_snapshot_ra_all} shows example temperature fields for the cases we study, at a snapshot in time after the flow has equilibriated. 
It also shows the different temperature profiles found in these cases (bottom), and compares them to the linear conductive state. 
We can see that as $\Ra$ increases, the profile becomes more uniform in the bulk, with a steeper temperature gradient in the boundary layers.


\section{Implementation}

\subsection{Parareal Algorithm}

The Parareal algorithm, first introduced in Lions et al. in 2001 \cite{lions2001resolution}, is briefly outlined here. 
For a more in depth explanation, see for example \cite{gander2007analysis}.

Parareal is a method used to speedup numerical solutions of initial value problems (IVPs) of the form 
\begin{equation}
    \frac{\partial U}{\partial t} = f\left(U(t),t\right), ~~ U(0)=U_0, ~~ 0\leq t \leq t_\text{end}.
\end{equation}
Parareal makes use of a coarse solver $\mathcal{G}$ and a fine solver $\mathcal{F}$. 
The time domain is split into $N$ time slices, where $N$ is the number of processors available for parallelisation in the time domain. 
The fine solver is the numerical method with properties designed to give the solution to the system to a required degree of accuracy. 
The coarse method is a cheaper method designed to give an answer quicker than the fine method, and with reduced accuracy. 
The Parareal method iterates over the fine and coarse solvers to improve the accuracy of the initial solution given by the coarse solver, until it is as accurate as the fine solver. 
This is done using the correction step
\begin{align}
\nonumber U_{n+1}^{k+1} &= \mathcal{G}(t_{n+1},t_n,U_n^{k+1})\\&+  \mathcal{F}(t_{n+1},t_n,U^k_n)-\mathcal{G}(t_{n+1},t_n,U_n^k),
\label{eq:parareal}
\end{align}
where $n$ denotes the current time slice, and $k$ denotes the Parareal iteration number. 
The coarse solver operates in serial, hence the need for a cheaper solution method, whilst the fine solver is able to operate in parallel, the key to reducing solution times.

\subsection{Spatial discretization}

We use a collocation-based pseudo-spectral method for the spatial discretisation, using Fourier bases with periodic boundaries for the horizontal ($x$) direction, and Chebyshev polynomial bases for the vertical ($z$) direction. 
The spatial resolution of a simulation is described by the number of collocation points in $x$ ($N_x$), and in $z$ ($N_z$). 
Simulations use the open source code Dedalus \cite{burns2019dedalus}, with the parareal\_dedalus \cite{clarke2019parareal} module used to implement the Parareal algorithm in the Dedalus solver. 
Time stepping is done using Implicit-Explicit Runge-Kutta timestepping methods (Ascher et al. \cite{ascher1997implicit}). Linear terms (diffusion, pressure and buoyancy forcing) are treated implicitly, whilst non-linear terms are treated explicitly. This combination lends itself to the pseudo-spectral method, as transformations between spectral and grid space are carried out using the parallel FFTW package, allowing multiplications to take place in grid space. 

\subsection{Validation}
\begin{figure}
    \centering
    \includegraphics{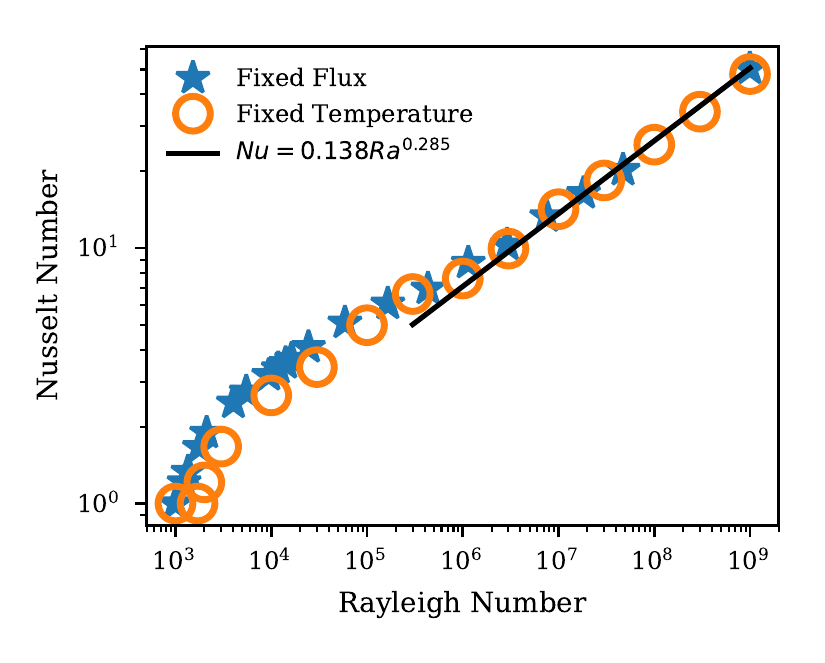}
    \caption{Calculated Nusselt values compared with the scaling found in Johnston and Doering \cite{johnston2009comparison}. Scaling of $0.135 \Ra^{0.286}$ was calculated from our data, compared to $0.138 \Ra^{0.285}$ found in \cite{johnston2009comparison}. Fixed temperature and fixed flux boundary simulations collapse on to the same line at high Rayleigh number, in agreement with \cite{johnston2009comparison} (black line).}
    \label{fig:nuselt_scaling}
\end{figure}

\begin{sloppypar}
The code was validated against the data in Johnston and Doering~\cite{johnston2009comparison}, see Figure~\ref{fig:nuselt_scaling}. 
Both fixed flux and fixed temperature boundary conditions were simulated.
We calculated a Rayleigh Nusselt scaling of $0.135 \Ra^{0.286}$ from our fixed flux data, very close to the $0.138 \Ra^{0.285}$ reported in \cite{johnston2009comparison}. 
The slightly higher Nusselt numbers that they found for fixed flux cases at low Rayleigh number were also replicated. 
Finally, we calculated the critical Rayleigh number by running multiple simulations near $Ra_c$ and checking the growth rate of the kinetic energy. 
We found that it was in agreement with Chandresakhar~\cite{chandrasekhar1961hydrodynamic} to within 0.1\%, 
\end{sloppypar}

\FloatBarrier

\subsection{Determining Accuracy of Fine Solution}
\label{sec:fine_accuracy}


We set a tolerance level of less than $1\%$ for $ \Nu_\text{int} $ defined in Equation \ref{eq:nu_int_error} and $ |P-\epsilon_U|/P $ defined in Equation \ref{eq:balance}. We also require a minimum of 6 points in the thermal boundary layers, that is $N_\text{BL} \geq 6$. 
At each $Ra$ we start with a low resolution ( ($N_x$, $N_z$) = (16, 8) for $\Ra=10^5$ and $10^6$ and $(32, 16)$ for $\Ra=10^7$) and then double the resolution in both spatial directions until all three conditions are met.

For comparison, we also carry out spatial convergence tests, comparing results obtained from the low resolution simulations with those obtained from a high resolution simulation for each $\Ra$ in the $L^2$ norm.
These are not used to determine the spatial resolution, though.
We calculate the relative difference in the final state temperature field by taking the $L^2$ norm with the high resolution (double resolution of shown values for each $\Ra$) final state. The second test is for $\Nu$, for which we calculate
\begin{equation}
    \Nu_\text{rel} = \frac{\left| \Nu - \Nu_\text{HR} \right|}{\Nu_\text{HR}},
\end{equation}
where HR denotes the high resolution simulation.


Table \ref{tab:resolution} shows the resolution required to meet the  consistency checks discussed above. 
We can see that the resolution required for 6 points in the boundary layer is higher than the resolution required for the other convergence tests, except for the $L^2$ error for $\Ra=10^7$. 
Figure~\ref{fig:spatial_convergence} shows how the $L^2$ error compares with $\Nu_\text{int}$. 
At $\Ra=10^5$, the resolution for a 1\% $L^2$ error is the same as what is required for the 1\% tolerance in the Nusselt numbers and buoyancy production and only half the resolution needed to have at least six nodes in the boundary layers. 
At $\Ra=10^7$, the $L^2$ error is not yet below 1\% even when all other tests are below tolerance, showing a significant difference in the $L^2$ error and the convergence tests we have set.
Given that the $L^2$ norm is not a very relevant quantity for understanding flow dynamics, if the internal checks and key quantities are converged before the $L^2$ error, then the lower resolution is deemed sufficient. 
The effect of timestep size on the accuracy of the solution was also investigated. 
However, it was found that for a given spatial resolution, the largest stable timestep was found to meet all of the accuracy criteria.

\FloatBarrier

\begin{figure*}[th]
\makebox[\linewidth][c]{%
\centering
\begin{subfigure}[b]{0.4\paperwidth}

\includegraphics[]{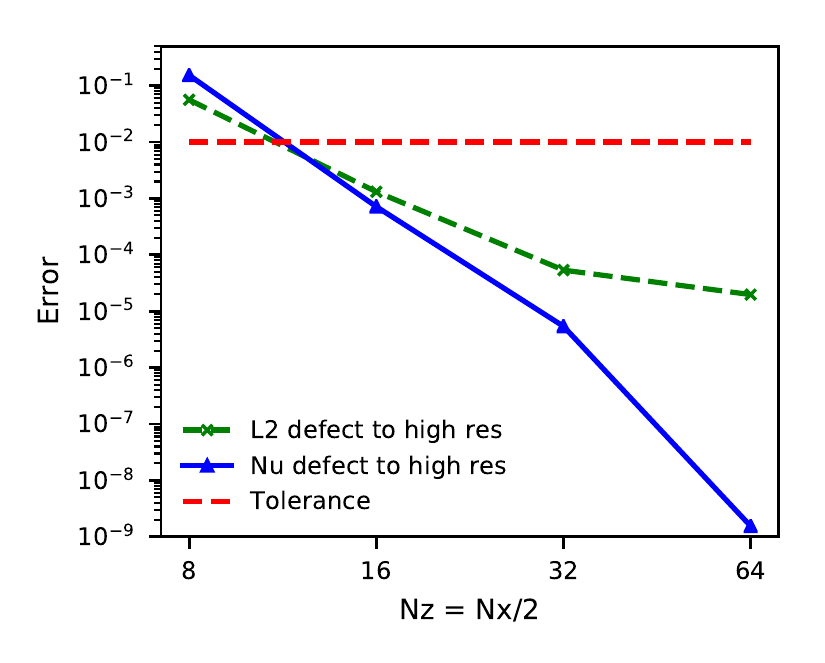}
\caption{$Ra=10^5$}
\end{subfigure}%
\begin{subfigure}[b]{0.4\paperwidth}
\includegraphics[]{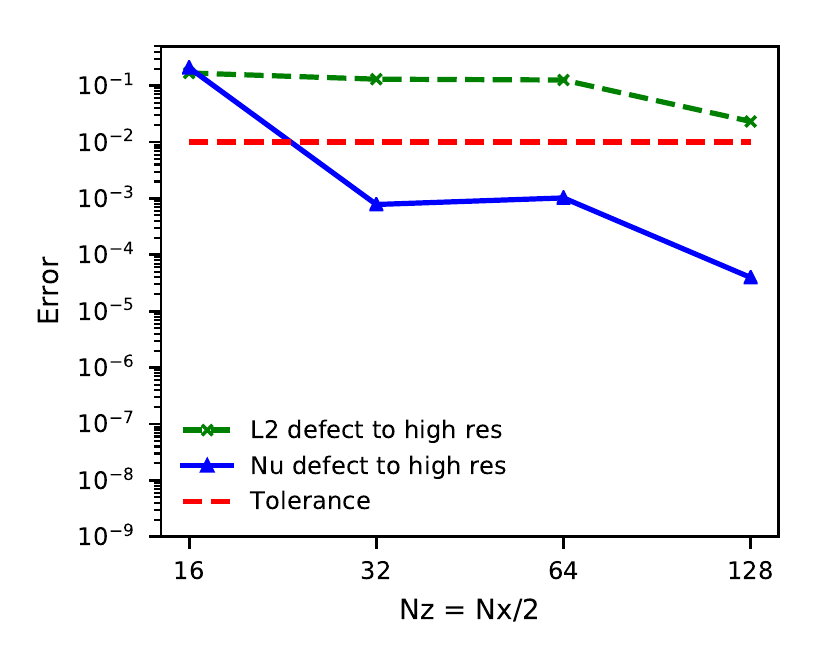}
\caption{$Ra=10^7$}
\end{subfigure}%
}\\
\caption{Spatial convergence of Nusselt number and $L^2$ errors relative to high resolution solution for $Ra=10^5$ (left) and $Ra=10^7$ (right). As expected, higher resolution is required for both quantities to meet the $10^{-2}$ tolerance for the higher Rayleigh number case. It can also be seen that the $L^2$ error requires much more resolution at higher Rayleigh number than the Nusselt number, where as at $Ra=10^5$, the resolution required to give good answers for the Nusselt number and $L^2$ error are similar. The shown Nusselt number is calculated by averaging over time and space.}
\label{fig:spatial_convergence}
\end{figure*}

\begin{table}[]
\caption{Table showing resolution required to meet various convergence tests. $L^2$, $\Nu_\text{int}$, $\Nu_\text{rel}$, and $|P-\epsilon_U|/P$ all have tolerance values of 1\%. $\Ra$ is the Rayleigh number, $N_\text{BL}$ denotes the resolution required for 6 points to be in the thermal boundary layer, $L^2$ denotes the defect of the end state temperature field to the high res simulation, $\Nu_\text{int}$ shows $\max (|\Nu_V-\Nu_b|,|\Nu_V-\Nu_t|,|\Nu_b-\Nu_t| ) / \Nu_V$, $\Nu_\text{rel}$ is the Nusselt number compared with the high resolution simulation, and $|P-\epsilon_U|/P$ is the buoyancy/ dissipation internal consistency check.}
\label{tab:resolution}
\begin{tabular}{ccllll}
\hline
$Ra$ & & \multicolumn{4}{c}{Resolution ($N_x, N_z$) for error $\leq 1\%$} \\
\multicolumn{1}{l}{} & \multicolumn{1}{l}{$N_\text{BL} \geq 6$} & $L^2$ & $\Nu_\text{int}$ & $\Nu_\text{rel}$ & $|P-\epsilon_U|/P $ \\ \hline
$10^5$ & (64,32) & (32,16) & (32,16) & (32,16) & (32,16) \\
$10^6$ & (128,64) & (64,32) & (64,32) & (64,32) & (32,16) \\
$10^7$ & (128,64) & (-,-) & (64,32) & (64,32) & (64,32)
\end{tabular}
\end{table}

\subsection{Duration of Simulation}
We determined the duration of a simulation based on a fixed number of advective times. There are three main timescales for Rayleigh-B\'enard flow which can be found from dimensional arguments; the thermal diffusive time\-scale, thermal advective time\-scale, and the viscous time\-scale. Here we ignore the viscous time\-scale, as we set $\Pra$ to 1. In the non-dimen\-sionali\-sation we have chosen, the diffusive and advective timescales are linked by $\tau_\text{advective} = Re \times \tau_\text{diffusive} $. Following Mound et al. \cite{mound_davies_2017}, we run our simulations for a set number (in this case 100) of advective times, after the initial transient has balanced out. However, in the $Ra=10^5$ case, we restrict the simulation to 1 diffusive time unit, since the solution is effectively steady state.

\subsection{Choice of Coarse Solver}

There are several options for choosing a coarse solver for Parareal.
These include a lower order timestepper, a larger timestep, reduced spatial resolution, reduced physics, or a different method of solving the equations. In this work, we reduce the spatial resolution and reduce the timestep. In tests, we used different levels of spatial coarsening to find the optimal amount for speedup. We tested coarsening factors (CF) of 2, 4, and 8, where $(N_x,N_z)$ of the coarse solver is equal to $1/\text{CF}$ $(N_x,N_z)$ of the fine method.
A coarsening factor of 2  did not lead to a speedup. Convergence was quick, but the runtime of the coarse solver was too close to the that of the fine solver. 
A coarsening factor of 4 worked better, allowing for quick convergence along with a significant difference in the cost of the fine/coarse solvers. 
A factor of 8 reduction showed slow convergence, and was not pursued further.

When choosing a coarse time step, we found situations where a Parareal simulation could be unstable even when a stable coarse solver was combined with a stable fine solver. 
This is likely due to the stability of Parareal itself, which has its own stability criterion, separate to the individual solvers~\cite{StaffRonquist2005}. 
This leads to lower speedups as we had to use smaller coarse time steps, making the coarse solver more costly.
We also investigated using lower order timesteppers for the coarse solver, along with the reduced resolution. 
However, as the stability region of Runge-Kutta tends to increase with the order, we found that reduced timestep sizes were required for lower order coarse solvers. 
This cancelled out any speed increase from reduced computation, thus the higher order timestepper RK443 was used in both the fine and coarse solver.

\subsection{Determining Convergence in Parareal}
The most simple and widely used check for convergence in Parareal is to monitor the defect between two consecutive iterate~\cite{samaddar2017temporal,aubanel2011scheduling,berry2012event}.
This has the benefit of being easy to implement, and can be done whilst running the simulation. 
However, as discussed in Section~\ref{sec:fine_accuracy}, using the $L^2$ can lead to substantial over-resolution of the problem if one is interested only in the averaged dynamics.
Therefore, the typical online Parareal convergence test is not suitable in this case. 
Since, at the moment, no termination criteria for averaged dynamics has been published, we perform a fixed number of Parareal iterations and assess convergence in post processing.
While useful for benchmarking, this is obviously not a reasonable approach for production runs.
Research into alternative and more application-orienteded termination criteria for Parareal therefore seems to be an area were further studies are urgently needed.

\FloatBarrier

\section{Results}

\FloatBarrier

\subsection{Kinetic energy in the Parareal solution}

\begin{figure*}[th]
\makebox[\linewidth][c]{%
\centering
\begin{subfigure}[b]{0.4\paperwidth}

\includegraphics[]{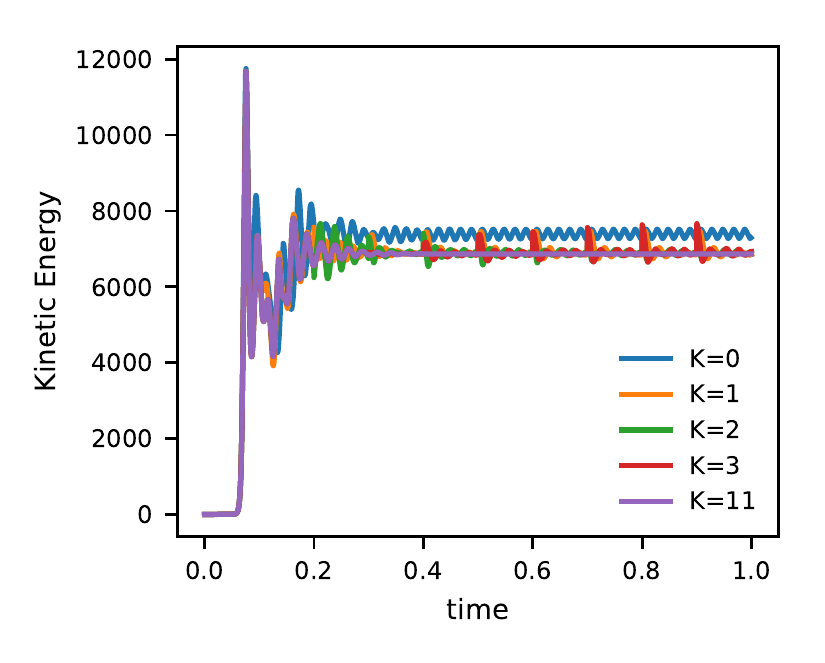}
\caption{$Ra=10^5$}
\label{fig:parareal_ke_time_ra_1e5}
\end{subfigure}%
\begin{subfigure}[b]{0.4\paperwidth}
\includegraphics[]{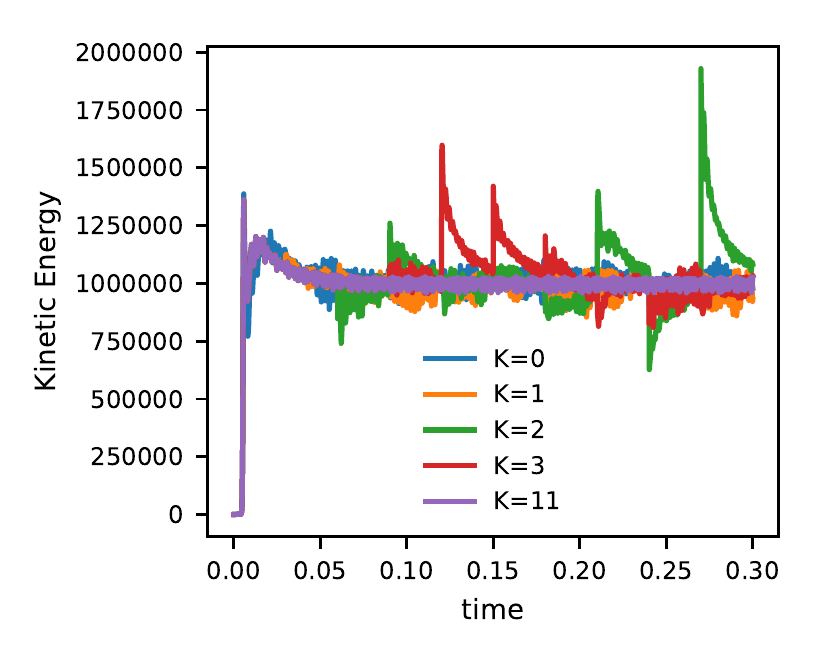}
\caption{$Ra=10^7$}
\label{fig:parareal_ke_time_re_1e7}
\end{subfigure}%
}\\
\caption{Kinetic energy against time for different numbers of Parareal iterations $K$ for $Ra=10^5$, $10^7$. The coarse solver has $\frac{1}{4}$ the number of modes in $x$ and $z$ as the fine solver, the coarse timestep is $\approx 2 \times$ the fine timestep, and the simulation used 10 time slices. The coarse solver for $Ra=10^5$ shows higher kinetic energy levels, along with periodic behaviour not present in the serial solution ($K=11$). For $10^7$, large jumps in the solution for $k>0$ are due to the Parareal correction step. The error at the jumps is growing, rather than shrinking, as the iteration number increases, showing the inability of Parareal to converge in this parameter regime. }
\label{fig:spatial_convergence}
\end{figure*}

Figures \ref{fig:parareal_ke_time_ra_1e5}, \ref{fig:parareal_ke_time_re_1e7} show the kinetic energy against time, for Rayleigh numbers $10^5$, $10^7$, for different numbers of Parareal iterations $k$. The number of time slices was kept constant at 10. For $Ra=10^5$, an initial Parareal coarse run shows significant differences from the subsequent Parareal iterations. The overall kinetic energy is higher in the low resolution coarse solver, and varies over time periodically. This increased kinetic energy in the coarse solver is due to dissipation of the system being under resolved at the coarse resolution. The periodicity is not present in the fine solution, and the effect can be seen to reduce in the subsequent iterations. The kinetic energy quickly reduces to the correct level after the first iteration for each time slice. Subsequent iterations still have a small 'bump' in kinetic energy at the correction time, but the overall level is in accordance with the fine solver. 
The $Ra=10^7$ case shows problems with the Parareal convergence. The correction steps increase the error, which can be seen in the large jumps at the time slice boundaries. This is the first indication that Parareal has reached the limit of usability in this parameter space.

\subsection{Parareal convergence}
\label{sec:parareal_convergence}

\begin{figure*}
\makebox[\linewidth][c]{%
\begin{subfigure}[b]{0.4\paperwidth}
\centering
\includegraphics[]{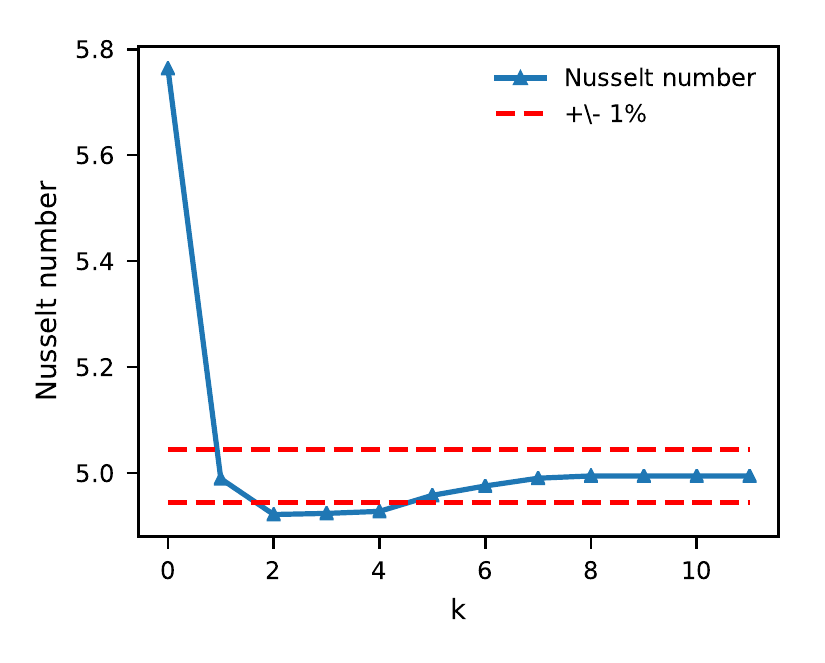}
\caption{$Ra=10^5$}
\end{subfigure}%
\begin{subfigure}[b]{0.4\paperwidth}
\includegraphics[]{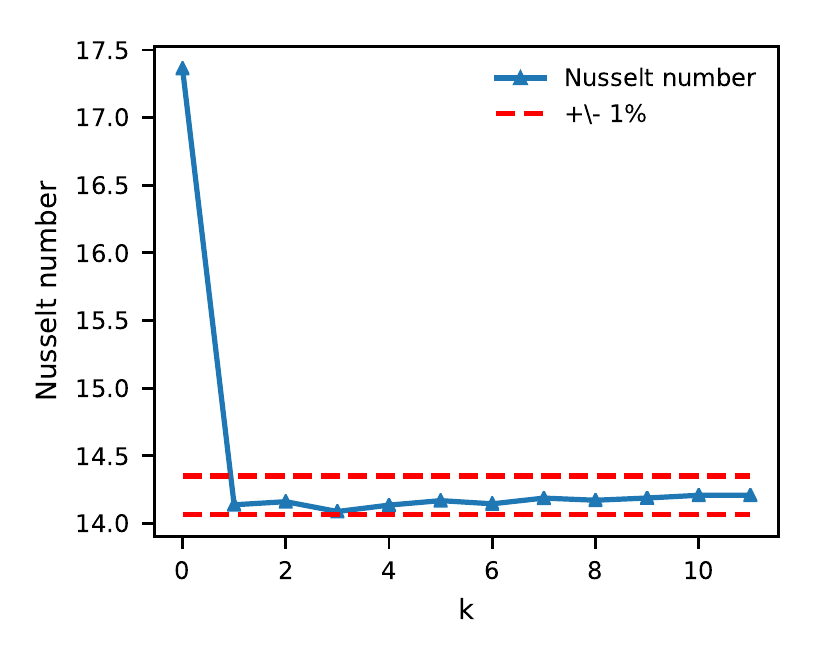}
\caption{$Ra=10^7$}
\end{subfigure}%
}\\
    \caption{Changing Nusselt number with Parareal iteration $k$. There is a large error in the Nusselt number calculated from the coarse solver ($k=0$), so that at least one iteration is required to calculate the correct Nusselt number (within 1\% - dotted red lines). For the Nusselt number alone, convergence behaviour is encouraging, for $Ra=10^5$ and $Ra=10^7$. The simulation was carried out with 10 time slices. }
    \label{fig:parareal_k_v_nusselt}
\end{figure*}

Figure \ref{fig:parareal_k_v_nusselt} shows how the calculated Nusselt number chan\-ges with increasing Parareal iterations. 
The Nusselt number found from the initial coarse solve is outside the accuracy requirement with an error of around 10\% rather than 1\%. 
In the case of  $Ra=10^5$, the Nusselt number converges to within the accuracy envelope after 1 iteration, but then in iterations 2-4 it falls back outside this region before converging again from iteration 5. 
For $Ra = 10^7$, the Nusselt number converges after a single iteration.

\begin{figure*}
\makebox[\linewidth][c]{%
\begin{subfigure}[b]{0.4\paperwidth}
\centering
\includegraphics[]{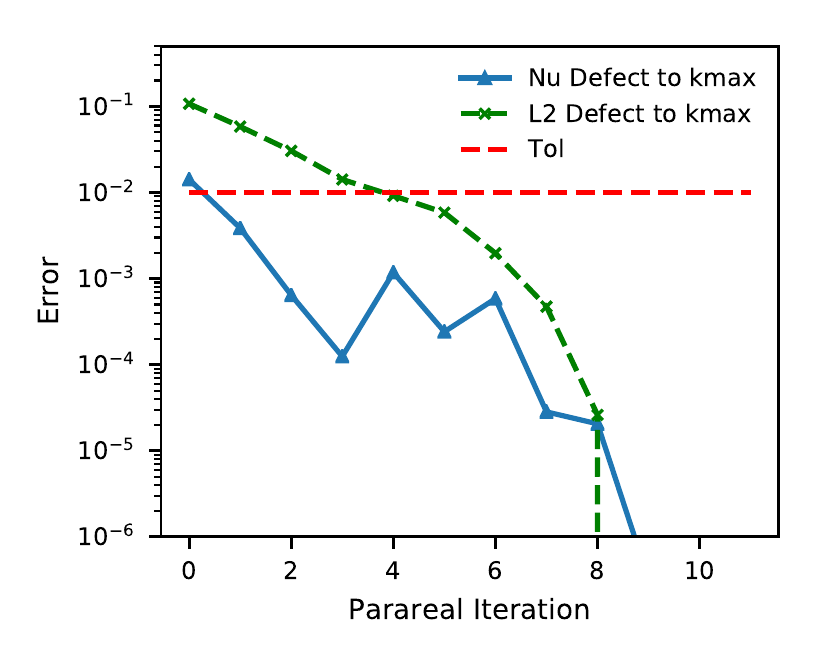}
\caption{$Ra=10^6$}
\end{subfigure}%
\begin{subfigure}[b]{0.4\paperwidth}
\includegraphics[]{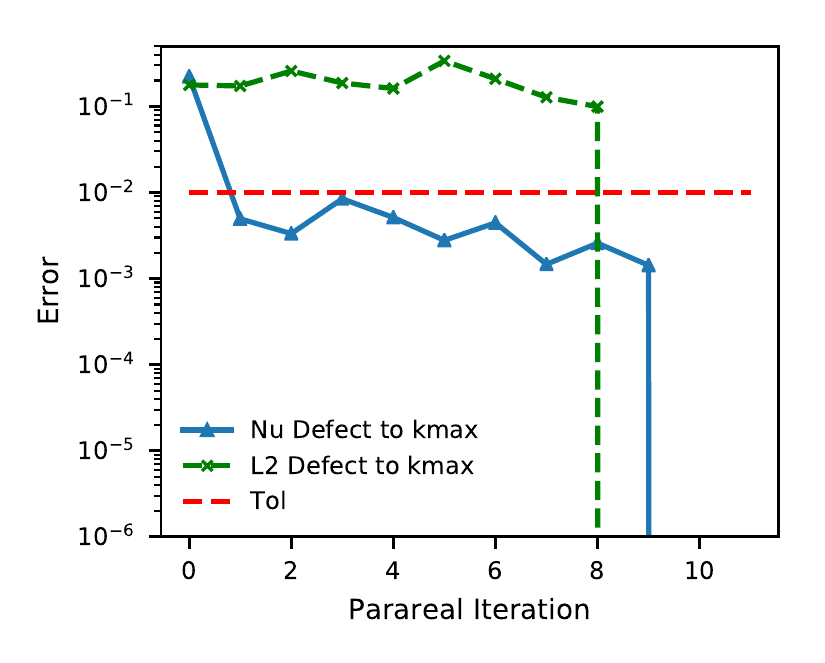}
\caption{$Ra=10^7$, 10 Time slices}
\end{subfigure}%
}\\
\begin{subfigure}{0.4\paperwidth}
\centering
\includegraphics[]{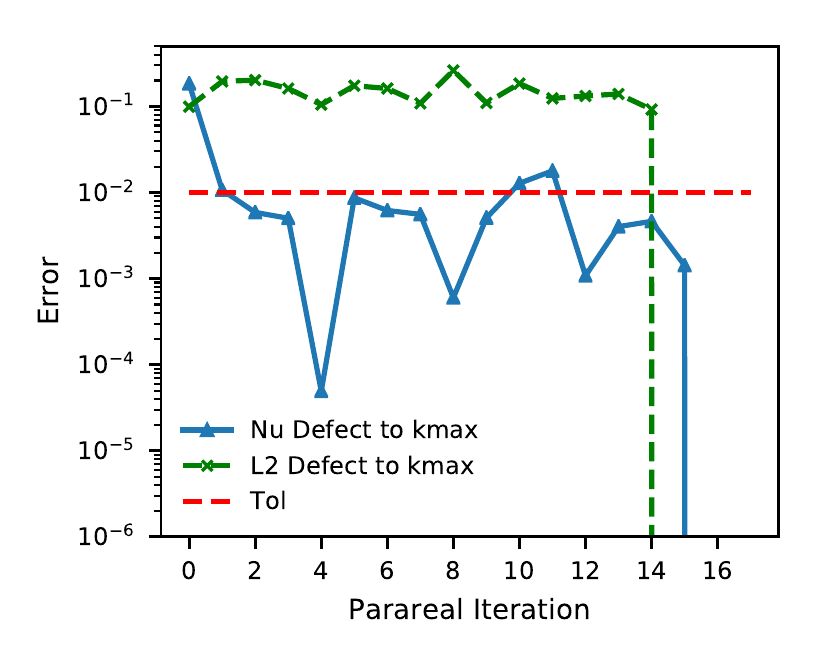}
\caption{$Ra=10^7$, 16 Time slices}
\end{subfigure}
    \caption{Convergence of Nusselt number and $L^2$ error with Parareal iteration for $Ra=10^6, 10^7$. As kmax is greater than number of timeslices, the solution at kmax perfectly represents the serial fine solution. We can see that the $L^2$ error at $Ra=10^6$ behaves as expected for good Parareal convergence, with a superlinear convergence behaviour. The Nusselt error at this Ra also shows convergence, but is more erratic. At $Ra=10^7$, we see much worse convergence. The $L^2$ error does not converge until the last iteration, when $k$ is equal to the number of time slices. The Nusselt number error behaves slightly better, but does not decrease monotonically. Figure(c) shows $Ra=10^7$ but with 16 time slices. Here, it requires two iterations for the Nusselt number to reach the $1\%$ tolerance.}
    \label{fig:parareal_L2_nu_error}
\end{figure*}

Figure~\ref{fig:parareal_L2_nu_error} shows the comparison of the $L^2$ error with the error in Nusselt number for $Ra=10^6$, $10^7$. 
In the smaller $Ra$ case, there is smooth convergence in both the $L^2$ error and in the Nusselt error, although the Nusselt convergence is slightly more erratic. 
In the $Ra=10^7$ case, we see that the Nusselt number error falls just underneath the tolerance threshold after the first iteration.
This is followed by a shallow decline in the error until the final iteration. 
The $L^2$ error behaves very differently, with a constant error of around 10\% right up until the $9^\text{th}$ iteration. 
We see here the mismatch in the error with respect to time averaged quantities with errors with respect to snapshots of the solution ($L^2$). 

\begin{figure*}
\makebox[\linewidth][c]{%
\begin{subfigure}[b]{0.4\paperwidth}
\centering
\includegraphics[]{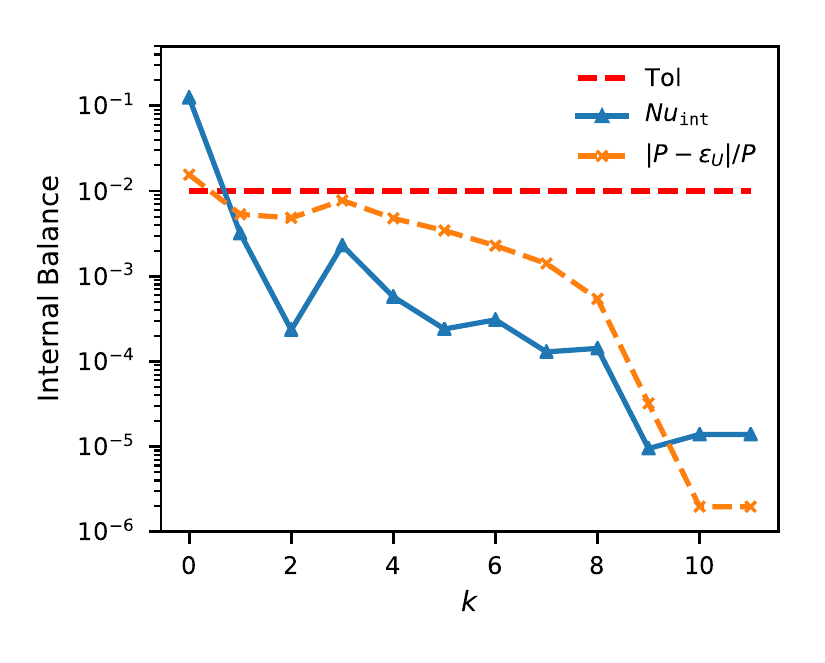}
\caption{$Ra=10^5$}
\end{subfigure}%
\begin{subfigure}[b]{0.4\paperwidth}
\includegraphics[]{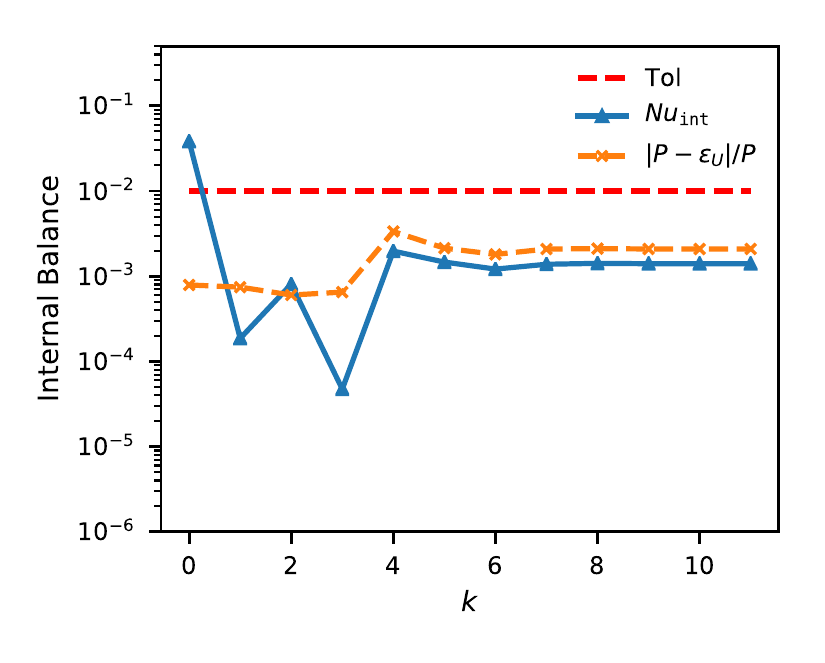}
\caption{$Ra=10^6$}
\end{subfigure}%
}\\
\begin{subfigure}{0.4\paperwidth}
\centering
\includegraphics[]{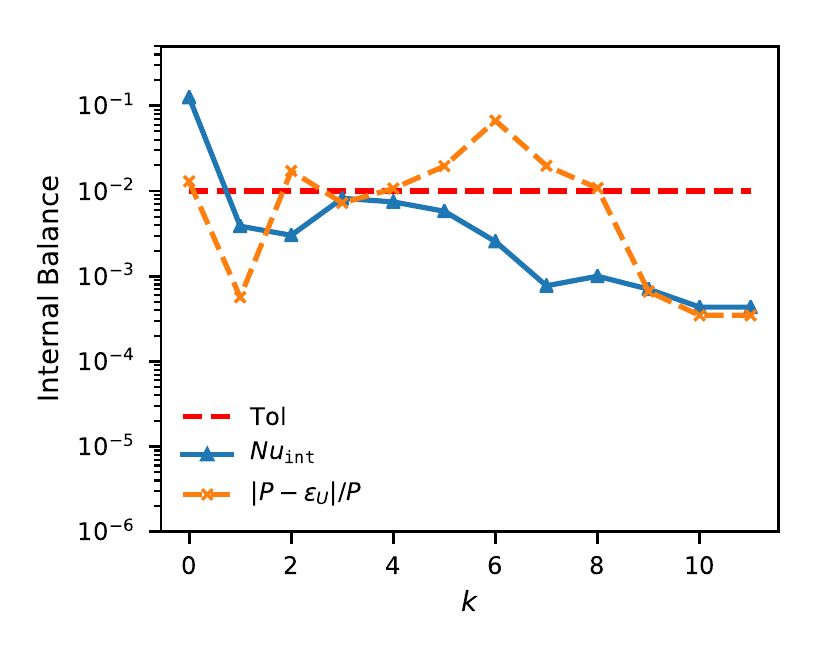}
\caption{$Ra=10^7$}
\end{subfigure}
    \caption{Convergence of the internal checks carried out on the data of the Parareal simulations, for $Ra=10^5, 10^6, 10^7$. The internal energy balance ($|P-\epsilon_U|/P$) takes longer to converge than $\Nu_\text{int}$. The Nusselt number is convergent for all three cases, but the internal energy balance is not convergent at the highest Rayleigh number.  }
    \label{fig:parareal_internal_errors}
\end{figure*}

 Figure \ref{fig:parareal_internal_errors} shows the internal consistency errors ($\Nu_\text{int}$, $|P-\epsilon_U|/P$) for all three $Ra$ tested. In all three cases, the $|P-\epsilon_U|/P$ and $\Nu_\text{int}$ converge to within the $1\%$ tolerance after one iteration. 
 However, the results for $Ra=10^7$ show that $|P-\epsilon_U|/P$ then returns above the tolerance level, and does not fall reliably until 8 iterations have been completed. 

We have also carried out numerical experiments for different numbers of time slices. 
Here, we would expect to see a trend where the number of iterations required to converge slowly increases with the number of time slices. 
In our results, we found that the number of iterations required did not behave like this for $Ra=10^7$. 
The number of iterations required increased and decreased with no clear pattern up to 20 time slices. 
Beyond this the iteration count was always higher than 1, and gradually increased with the number of time slices.

\subsection{Scaling and Performance}

\begin{figure}
    \centering
    \includegraphics{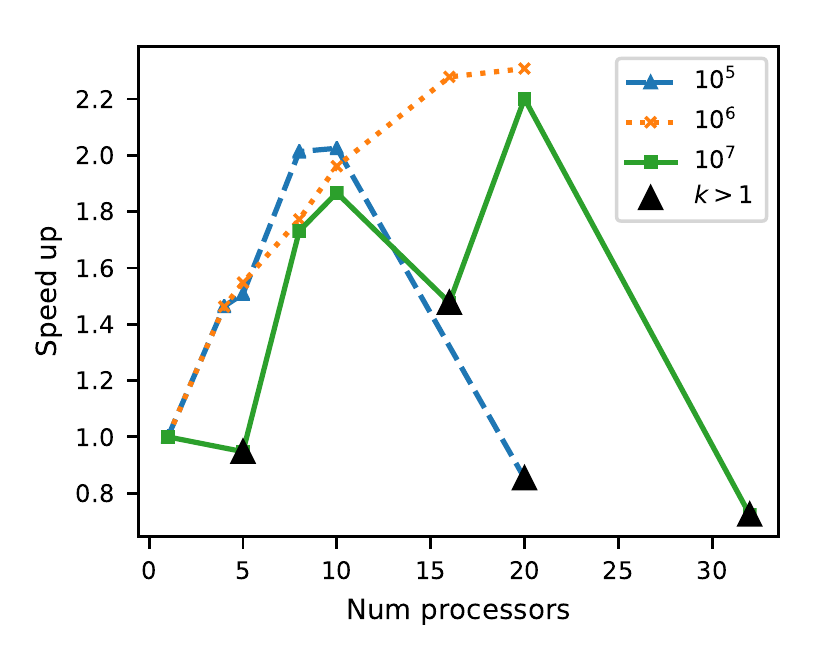}
    \caption{speedup vs number of timeslices/processors for $Ra=10^5$, $10^6$, and $10^7$. We can see that performance apeears best for $Ra=10^6$. Peak speedup is around 2 $\sim$ 2.4 for all $Ra$. For $10^5$, and $10^6$, performance is predictable, with speedup increasing with number of cores until a scaling limit is reached. For $10^7$, the scaling behaviour is erratic, due to the errors being very close to the tolerance limit. This leads to more iterations being required for convergence at some processor counts, causing the smaller speedups (black triangles).}
    \label{fig:parareal_scaling}
\end{figure}

Figure \ref{fig:parareal_scaling} shows the scaling performance for simulations with $Ra=10^5,~10^6,~10^7$. 
We see standard scaling behaviour for both $10^5$, and $10^6$, where speedup increases with processor count until the scaling limit is reached, and no further performance gains are possible. 
This is due to an increase in the number of Parareal iterations required at higher time slice count. 
We also see that performance is better at $10^6$ than at $10^5$, likely because the bigger problem size due to higher resolutions improves scaling.
However, the performance of Parareal at $Ra=10^7$ is much more mixed. This is in part due to the errors being very close to the tolerance level for all iterations after $k=1$, see Figure~\ref{fig:parareal_L2_nu_error}b. 
The error does not fall with increasing iterations in the way it does for $Ra=10^5$, $10^6$, rather, it hovers very close to the tolerance value.
Convergence behaviour with number of time slices is unpredictable in this case. 
For some numbers of time slices, such as in Figure~\ref{fig:parareal_L2_nu_error}b, the Nusselt error falls below tolerance after one iteration and remains there. 
In other cases, such as five or 16 time slices, see Figure~\ref{fig:parareal_L2_nu_error}c, the error falls below the tolerance and then rises back again. 


\section{Conclusions}\label{sec:conclusions}

\subsection{Parareal for Rayleigh-B\'enard convection}

We have shown that the Parareal algorithm allows for reliable speedup of simulations in a limited range of Rayleigh numbers at finite Prandtl number.
The algorithm converges quickly with respect to averaged quantities like the Nusselt number and internal energy balance.
Although slower, Parareal also converges with respect to the $L^2$ defect between subsequent iterations.
Speedups of up to 2.4 are possible, with around 20 processors, with parallel efficiencies of around 0.2 for Rayleigh numbers as high as $10^6$. 
However, in all cases, speedups were limited to at most 20 processors. Beyond that, increases in the number of required iteration balanced out any gains from using more processors.

At $Ra=10^7$, we find that convergence of Parareal degrades substantially.
The errors in $\Nu$ do not fall monotonically with increasing iteration number. 
For some simulations, the error falls below the tolerance level at a low number of iterations, only to increase in successive iterations. 
This erratic behaviour leads to irregular scaling performance at $10^7$; sometimes the simulation converges in one iteration, sometimes it takes two or three. 
These findings are in contrast to what Samuel~\cite{samuel2012time} found for $Ra = 10^7$ with infinite Prandtl number where he observed a small number of iterations independent of the number of time slices being required for convergence and increasing speedup up to 40 processors.
Clearly, performance of Parareal is very different in the finite versus infinite Prandtl number case.

Parareal is not expected to be useful for simulations of Rayleigh-B\'enard convection at Rayleigh numbers above $10^7$ as we expect the performance to degrade further as the flow becomes more turbulent. 
This difference in performance is caused in part by the well known general degradation of Parareal with increasing Reynolds numbers~\cite{steiner2015convergence}. 
It is also caused by the choice of convergence criteria. 
The correction step of Parareal depends on pointwise amplitude corrections at the boundary between time slices. 
In Rayleigh-B\'enard convection studies, the particular state of a given field at an instant in time is not of primary concern, therefore we relaxed the accuracy conditions of the fine solution, so that we did not enforce that the $L^2$ error be below a threshold value. 
In the cases of $10^5$ and $10^6$, the $L^2$ error is of roughly the same magnitude as the time- and space- averaged quantities ($\Nu_\text{int}$, $|P-\epsilon_U|/P$), used to determine accuracy of the solution. 
In the $10^7$ simulations, we can find a good level of accuracy in the $\Nu_\text{int}$ and $|P-\epsilon_U|/P$, whilst the $L^2$ error is still high in spatial convergence tests, (see Figure \ref{fig:spatial_convergence}). 
As the Parareal algorithm effectively operates on the $L^2$ error, Parareal convergence is slow. 
Exploring the performance of of other parallel-in-time methods like PFASST or MGRIT, and potentially a comparison with Parareal, would be an interesting direction for future research.

\subsection{Convergence of Statistical Quantities in Parareal}
For larger Rayleigh numbers, our tests show a significant disparity between the instantaneous $L^2$ error in a variable field such as temperature and the error in statistically calculated quantities such as the Nusselt number. 
In one example, Parareal reached a 1\% error with respect to the Nusselt number while the $L^2$ error stalled for 7 iterations and only fell below 1\% after iteration 8.
In a case like Rayleigh-B\'enard convection, statistical quantities like the Nusselt number are typically the most informative for understanding the behaviour of the physical system and what domain scientists are interested in. 
Therefore, we argue that this should be the criteria for determination of convergence, similar to what is used in time serial studies. 
However, for a reliable estimate of this kind of quantity, a time average is required across multiple time slices, in addition to a spatial average.  
Obtaining this kind of data during a simulation to monitor and terminate Parareal's convergence with respect to statistical quantities is a problem that presents an interesting challenge, and would be a useful avenue for further investigation.


%
%

\bibliographystyle{spmpsci}      
\bibliography{main}   

%
%

\end{document}